\documentstyle[editedvolume]{crckapb} 

\input psfig.tex

\newcommand{\be}{\begin{equation}}
\newcommand{\ee}{\end{equation}}
\def\lsim{\lower.5ex\hbox{$\; \buildrel < \over \sim \;$}}
\def\gsim{\lower.5ex\hbox{$\; \buildrel > \over \sim \;$}}

\begin{opening}
\title{Nucleosynthesis in Advective Accretion Disks Around Galactic and Extra-Galactic Black Holes}

\author{B. Mukhopadhyay}
\institute{S. N. Bose National Centre For Basic Sciences
	JD Block, Salt Lake, Sector-III, Calcutta-700091,
	India}

\end{opening}

\runningtitle{Nucleosynthesis in advective disk }

\begin{document}

\noindent Appearing in `Observational Evidence for Black Holes in the Universe',
Ed. S.K. Chakrabarti  (Kluwer Academic:Holland) p. 105 (1998).

\section{Introduction}

Many of the observational evidences for black hole rely on the
fact that the incoming gas has the potential 
to become as hot as its virial temperature $T_{virial} \sim 10^{13}$ $^oK$ 
(Rees, 1984). This flow is usually cooled down through bremsstrahlung 
and Comptonization effects and hard and soft states are produced depending 
on the degree by which this cooling takes place (Chakrabarti \& Titarchuk, 1995).
The generally sub-Keplerian, advective flow after deviating from a Keplerian disk,
especially in the hard states, remains sufficiently hot to cause a 
significant amount of nuclear reactions around a black hole before plunging 
in it. The energy generated could be high enough to destabilize the 
flow and the modified composition may be dispersed through winds to change 
the metalicity of the galaxy (Chakrabarti, Jin \& Arnett, 1987 [CJA]; Jin, 
Arnett \& Chakrabarti, 1988; Chakrabarti, 1988; Mukhopadhyay \& Chakrabarti, 
1998). Earlier works have been done in cooler thick accretion disks only. 
Below, we present a few examples of nuclear reactions in advective 
flows and discuss the implications. Results of more detailed study 
could be seen in Mukhopadhyay \& Chakrabarti (1998) [MC98].

\section{Physical Systems Under Considerations} 

Black hole accretion is by definition advective, i.e., matter must
have {\it radial} motion, and transonic, i.e., matter must be supersonic
(Chakrabarti 1996 [C96]
and references therein). The supersonic flow must be sub-Keplerian
and therefore deviate from the Keplerian disk away from the black hole.
The study of viscous, transonic flows was initiated by 
Paczy\'nski \& Bisnovatyi-Kogan (1981).

By and large, we follow C96 for thermodynamical
parameters along a flow and  Chakrabarti \&
Titarchuk (1995) [CT95] and Chakrabarti (1997a) [C97a] 
to compute the temperature of the Comptonized flow in the
advective region which may or may not have shocks.
According to these solutions, a black hole accretion 
may be thought to be similar to a sandwich whose
sub-Keplerian flow rate (${\dot m}_{h}$) in the `bread' part progressively
increases and that (${\dot m}_{d}$) in the `meat' part
progressively decreases as flow moves in towards the
black hole. Finally at $x=x_K$, the equatorial flow
also deviates from a Keplerian disk and for $x<x_K$
the entire flow is sub-Keplerian. Among the
major reactions which are taking place inside the disk, 
we note that, due to hotter nature of the advective disks,
especially when the accretion rate is low and Compton cooling
is negligible, the major process of
hydrogen burning is the rapid proton capture process (which operates
at $T \gsim 5 \times 10^8$K) as opposed to the PP chain (which
operates at much lower temperature $T \sim 0.01-0.2 \times 10^9$K)
and CNO (which operates at $T \sim 0.02-0.5 \times 10^9$K). 
The present paper being exploratory in nature, we do not include nuclear 
heating and cooling in determining the structure and stability
of the accretion flow. We do not assume here heating due to
magnetic dissipation (see, Shapiro, 1973 and Bisnovatyi-Kogan, 1998).

For simplicity, we  take the solar abundance as the abundance of the Keplerian disk.
Furthermore, Keplerian disk being cooler, no composition change is assumed inside
it.  In other words, our computation starts only from the
time when matter is launched from the Keplerian disk ($x=x_K$).
Most of the cases were repeated with initial abundance same as 
the output of big-bang nucleosynthesis (hereafter referred to as
`big-bang abundance').

According to CT95, and C97a, for two component
accretion flows, for ${\dot m}_d$$\lsim 0.1$ and ${\dot m}_h$$\lsim 1$ the black
hole remains in hard states. Lower rate in Keplerian disks {\it generally}
implies a lower viscosity and a larger $x_K$ ($x_K \sim 30-1000$; see,
C96 and C97a). In this parameter range the protons remain hot,
typically, $T_p \sim 1-10 \times 10^9$ degrees or so. This is because the
efficiency of emission is lower ($f=1-Q^-/Q^+\sim 0.1$, where,
$Q^+$ and $Q^-$ are the heat generation [due to viscous processes] and heat loss rates
respectively. Also see, Rees [1984], where it is argued that ${\dot m}/\alpha^2$ is a
good indication of the cooling efficiency of the hot flow.). We have
studied a large region of parameter space in details where $0.0001 \lsim
\alpha \lsim 1$, $0.001\lsim {\dot m} \lsim 100$, $0.01 \lsim F_{Compt}
\lsim 0.95$, $4/3 \lsim \gamma \lsim 5/3$ are chosen. Here, $F_{Compt}$ is the
factor by which the proton temperature is reduced due to bremsstrahlung and Comptonization effects.
Results with several sets of initial conditions are in MC98. Since shocks can 
form in advective disks for a large region of parameter space (C96 and references therein)
we use a case with a standing shock in this paper.

In selecting the reaction network we kept in mind the fact that hotter
flows may produce heavier elements through triple-$\alpha$ and rapid
proton and $\alpha$ capture processes.Furthermore due to photo-dissociation
significant neutrons may be produced and there is a possibility of
production of neutron rich isotopes. Thus, we consider sufficient number
of isotopes on either side of the stability line. The network thus
contains protons, neutrons, till $^{72}Ge$ -- altogether 255 nuclear
species. The standard reaction rates were taken [MC98].

\section{Results}

We present now with a typical case which contained a shock wave
in the advective region. We use the  mass of the black hole
$M/M_\odot=10$, $\Pi$-stress viscosity parameter $\alpha_\Pi=0.07$,
the location of the inner sonic point $x_{in}=2.9115$ and the value of the specific angular
momentum at that point $\lambda_{in}=1.6$, the polytropic index $\gamma=4/3$ 
as free parameters. The net accretion rate ${\dot m}=1$, 
which is the sum of (very low)  Keplerian component and the sub-Keplerian 
component. Results of  CT95 and C97a for ${\dot m}_d \sim 0.1$ and ${\dot m}_h \sim 0.9$,
fix $F_{Compt} = 0.03$, $x_K=401$. This factor is used to convert the
temperature distribution of solutions of C96 (which does not
explicitly uses Comptonization) to temperature distribution {\it with}
Comptonization. The proton temperature  and velocity distribution computed in this
manner are shown in Figs. 1(a-b).
(velocity is measured in units of $10^{10}$ cm sec$^{-1}$).

\begin{figure}
\vbox{
\vskip -0.5cm
\hskip -11.5cm
\centerline{
\psfig{figure=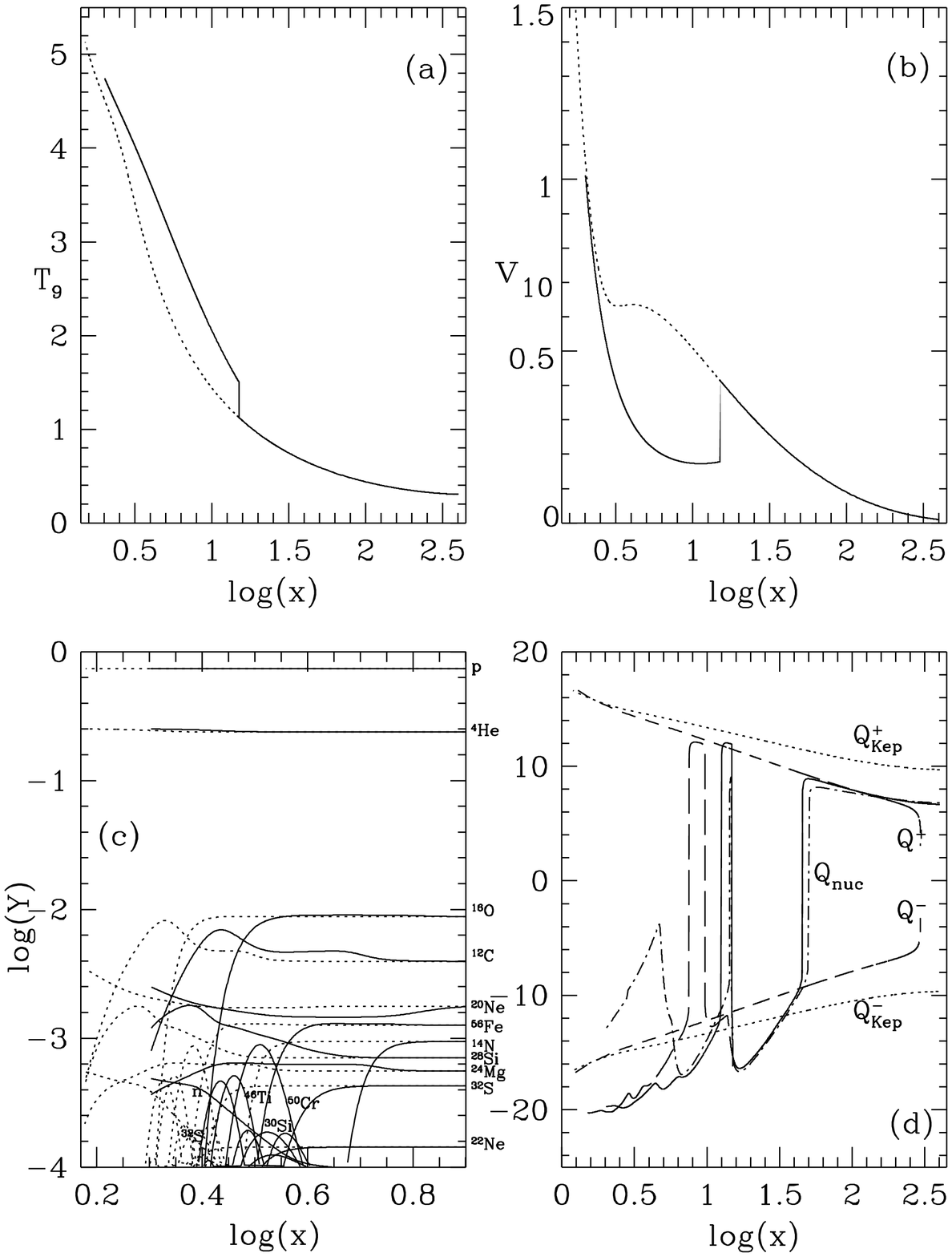,height=13truecm,width=13truecm,angle=0}}}
\vspace{-0.0cm}
\noindent {\small {\bf Fig. 1} : Variation of (a) proton temperature ($T_9$), (b) radial velocity $v_{10}$,
(c)  matter abundance $Y_i$ in logarithmic scale and (d) various forms of specific energy release
and absorption rates as functions of logarithmic radial distance ($x$ in units of Schwarzschild radius).
See text for parameters. Solutions in the stable branch with shocks are solid curves 
and those without the shock are dotted in (a-c). Curves in (d) are described in the text.
At the shock temperature and density rise significantly and cause
a significant change in abundance even farther out. Shock induced winds may
cause substantial contamination of the galactic composition when parameters are
chosen from these regions. }  
\end{figure}

In Fig. 1c, we show the composition change close to the black hole
both for the shock-free branch (dotted curves) and the shocked branch
of the solution (solid curves). Only prominent elements are plotted.
The difference between the shocked and the shock-free cases is that in the shock
case the similar burning takes place farther away from the black hole
because of much higher temperature in the post-shock region.
A significant amount of the neutron
(with a final abundance of $Y_n \sim 10^{-3}$) is produced due
to photo-dissociation process. Note that closer to the black hole,
$^{12}\!C$, $^{16}\!O$, $^{24}\!Mg$ and $^{28}\!Si$ are all destroyed
completely, even though at around $r=3$ or so, the abundance of some of them
went up first before going down. Among the new species which are formed
closer to the black hole are $^{30}\!Si$, $^{46}\!Ti$, $^{50}\!Cr$. Note that
the final abundance of $^{20}\!Ne$ is significantly higher than the initial value. Thus a significant
metalicity could be supplied by winds from the centrifugal barrier.  
In Fig. 1d, we show all the energy release/absorption components for the shocked flow. 
The viscous energy generation ($Q^+$) and the loss of energy ($Q_-$) from the disk (short dashed)
are shown. These quantities, had the advective regime had Keplerian
distribution, are also plotted (dotted). Solid curve represents the nuclear energy release/absorption
for the shocked flow and the long dashed  curve is that for the shock-free flow. 
Dot-dashed curve represents the nuclear energy release/absorption for big-bang abundance.
As matter leaves the Keplerian flow, the rapid proton capture ($rp$-) 
processes (such as, $p+ ^{18}\!O \rightarrow  ^{15}\!N + ^{4}\!He$ etc.)
burn hydrogen and releases energy to the disk.
At around $x=45$, $D \rightarrow n + p $ dissociates $D$
and the endothermic reaction causes the nuclear energy release to become 
`negative', i.e., a huge amount of energy is absorbed from the disk. 
At around $x=14$ the energy release is again dominated by the  original {\it rp}-processes.
Excessive temperature at around $x=12.6$ breaks $^3\!He$ down into deuterium
This type of reactions absorb a significant amount of energy from the flow. 
When big-bang abundance is chosen to be the initial abundance, the net composition does
not change very much, but the dominating reactions themselves are
somewhat different because the initial compositions are different.
For instance, in place of rapid proton capture
reactions as above, the fusion of deuterium into $^4\!He$ plays dominant role via
$D+D \rightarrow ^3\!He+n$, $D+p \rightarrow ^3\!He$, $D+D \rightarrow 
p+T$, $^3\!He+D \rightarrow p+^4\!He$. This is because no heavy elements were present
to begin with. Endothermic reactions at around $x=20-40$ are dominated by 
deuterium dissociation as before. However,
after the complete destruction of deuterium, the exothermic reaction is
momentarily dominated by neutron capture processes 
(due to the same neutrons which are produced earlier via $D \rightarrow n+p$) 
such as $n+^3\!He \rightarrow p +T$ which produces the spike at 
around $x=14.5$. Following this, $^3 \!He$ and $T$ are destroyed
as in solar abundance case and reaches the minimum
in the energy release curve at around $x=6$. The tendency of going back to the
exothermic region is stopped due to the photo-dissociation of $^4\!He$ via
$^4\!He \rightarrow p+T$ and $^4\!He \rightarrow n + ^3 \!He$. At the end 
of the big-bang abundance calculation, a significant amount of neutrons are 
produced.  It is interesting to note that the radial dependence
as well as the magnitude of the energy release due to $rp$-process and
that due to viscous dissipation ($Q^+$) are {\it very} similar
(save the region where endothermic reactions dominate). 
This suggests that even with nuclear reactions, 
at least some part of the advective disk may be perfectly stable.

\begin{figure}
\vbox{
\vskip -5.0cm
\hskip -9.0cm
\centerline{
\psfig{figure=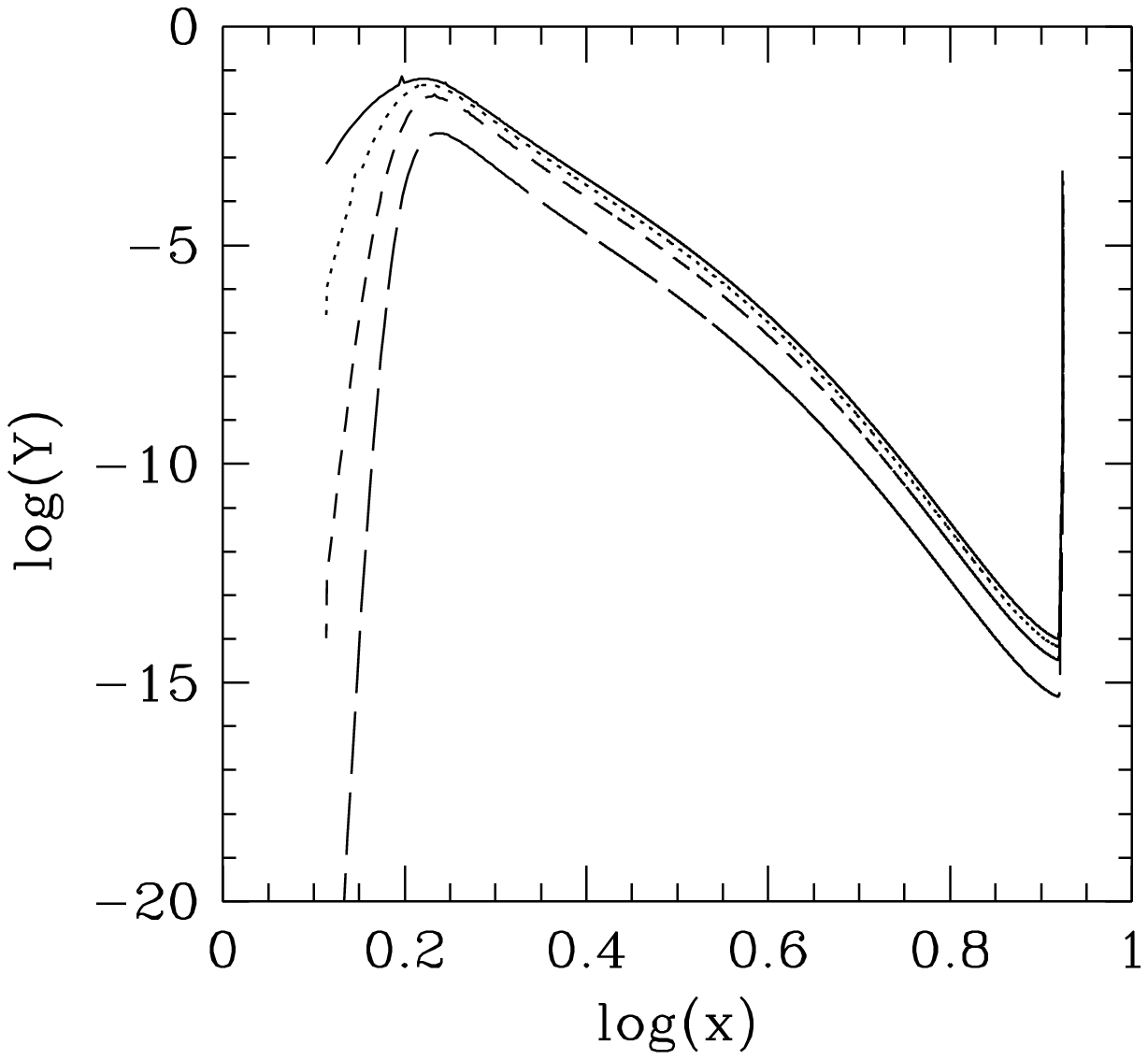,height=12truecm,width=12truecm,angle=0}}}
\vspace{-0.0cm}
\noindent {\small {\bf Fig. 2 :}
The convergence of the neutron abundance through successive iterations in a
very hot advective disk. From bottom to top curves $1$, $7$, $14$ and $21$
iteration results are shown. A neutron torus with a significant abundance is
formed in this case. }
\end{figure}

We now present another interesting case where lower accretion rate
(${\dot m}=0.01$) but higher viscosity ($0.2$) were used and the efficiency of emission
is intermediate ($f=0.2$). That means that the temperature of the flow is high
($F_{Compt} = 0.1$, maximum temperature $T_9^{max}=13$). $x_K=8.4$ in this case, 
if the high viscosity is due to stochastic
magnetic field, protons would be drifted towards the black hole
due to magnetic viscosity, but the neutrons will not be
drifted (Rees et al., 1982) till they decay. This principle has been used
to do the simulation in this case. The modified composition in one
sweep is allowed to interact with freshly
accreting matter with the understanding that the accumulated neutrons
do not drift radially. After few iterations or sweeps the steady
distribution of the composition may be achieved. Figure 2
shows the neutron distributions in iteration numbers
$1$, $7$, $14$ \& $21$ respectively (from bottom to top curves)
in the advective region. The formation
of a `neutron torus' (Hogan \& Applegate, 1987) is very apparent in this result and generally in all
the hot advective flows. Details are in Chakrabarti \& Mokhopadhyay (1998).

\section{Discussions and Conclusions}

In this paper, we have explored the possibility of nuclear reactions in
advective accretion flows around black holes. Although this region 
is not fully self-consistently computed yet, particularly near 
the region where the advective disk joins with a standard 
Keplerian disk, we have used the best model that is available in the
literature so far (C96). Temperature in this region is controlled by the 
efficiencies of bremsstrahlung and Comptonization processes (CT96, C97a) and 
possible heating and cooling due to magnetic fields (Shapiro, 1973; Bisnovatyi-Kogan, 1998). 
For a higher Keplerian rate and higher viscosity, the inner edge of the Keplerian
component comes closer to the black hole and the advective region becomes
cooler (CT95). However, as the viscosity is decreased, the inner edge of the Keplerian 
component moves away and the Compton cooling becomes less efficient.

The composition changes especially in the centrifugal pressure supported denser region,
where matter is hotter and slowly moving. Since centrifugal 
pressure supported region can be treated as an effective
surface of the black hole which may generate winds and outflows in the
same way as the stellar surface, one could envisage that the winds 
produced in this region would carry away modified composition
(Chakrabarti, 1997b; Das \& Chakrabarti 1998; Das, 1998). 
In very hot disks, a significant amount of 
free neutrons are produced which, while coming out through winds may 
recombine with outflowing protons at a cooler environment
to possibly form deuteriums a process originally suggested by Ramadurai \& Rees
(1985) in the context of ion tori around black holes. A few related 
questions have been asked lately: Can lithium in the universe be 
produced in black hole accretion (Jin 1990; Yi \& Narayan, 1997)?
We believe that this is not possible. The spallation reactions may produce 
such elements when only He-He reactions are considered. But when the
full network is used we find that the hotter disks where 
spallation would have been important also photo-dissociate 
heliums to deuteriums and then to protons and neutrons before any significant
lithiums could be produced. Another question is: Could the metalicity
of the galaxy be explained, at least partially, by nuclear reactions?
We believe that this is quite possible. Details are in MC98.

An interesting possibility of formation of the neutron torus was
also discussed by Hogan \& Applegate (1987): Can a neutron torus
be formed around a black hole?  We find that in the case of hot inflows,
such formation of neutron tori is a very distinct possibility (Chakrabarti
\& Mukhopadhyay, 1998). Presence of a neutron torus around a black hole
would help the formation of neutron rich species as well, a process hitherto 
attributed to the supernovae explosions only.

The advective disks as we know today do not perfectly match with a Keplerian disk. 
The shear, i.e., $d\Omega/dx$ is always very small in the advective 
flow compared to that of a Keplerian disk near the outer boundary of the 
advective region. We believe that such behavior is
unphysical and had the viscosity $\alpha$ parameter or the cooling
function were allowed to be  changed continuously, such deviation
would not have occurred. Thus some improvements of the disk model at the
transition region is needed, but since major reactions are closer to the
black hole, we believe that such modifications of the model would not change
our conclusions. The neutrino luminosity is generally very 
large compared to the photon luminosity in case of hot disk
(Mukhopadhyay \& Chakrabarti 1998). In the first Case that we 
discussed above, neutrinos typically carry an energy of around $10^{30}$ 
ergs sec$^{-1}$ gm$^{-1}$. Assuming that a typical neutrino is of energy $\sim 1$ MeV, and appreciable
neutrinos are emitted only from a region of a radial extent
of the order of a Schwarzschild radius where the disk is also around a 
Schwarzschild radius thick and the density is around $10^{-9}$ gm sec$^{-1}$.
In presence of hot advective disks, the number of neutrinos that should be
detected per square cm area on the surface of earth would be at least a
few per second provided the source is a $10M_\odot$ 
black hole at a distance of $10$kpc. On the other hand,
neutrino luminosity from a cool advective disk is low
(around $10^{15}$ ergs sec$^{-1}$ gm$^{-1}$) and no appreciable number of
neutrino are expected. Thus, probably one way to check if hot, and stable 
advective disks exist is to look for neutrinos from the suspected black 
hole candidates, especially in the hard states.

In all the cases, even when the nuclear composition changes are not
very significant, we note that the nuclear energy release due to exothermic
reactions or absorption of energy due to endothermic reactions is
of the same order as  actual radiation from the disk.
Unlike the gravitational energy release due to viscous processes,
nuclear energy release strongly depends on temperatures.
Thus, the additional energy source or sink may destabilize
the flow. This aspect has not been studied in this work yet. A realistic
way to do this is to include the nuclear energy also in time dependent 
studies of the black hole accretion (e.g., Molteni, Lanzafame \& Chakrabarti, 
1994; Molteni, Ryu \& Chakrabarti, 1997). Such works are in progress and the 
results would be reported elsewhere.

\end{document}